\documentclass[conference]{IEEEtran}
\usepackage{slashbox}
\usepackage{graphicx, epstopdf}
\usepackage{amsmath}
\usepackage{amsfonts}
\usepackage{amssymb}

\renewcommand\IEEEkeywordsname{Keywords}

\begin{document}

\title{Analysing the role of entanglement in the three-qubit Vaidman's game}

\author{\IEEEauthorblockN{Hargeet Kaur}
\IEEEauthorblockA{Department of Chemistry\\
Indian Institute of Technology\\
Jodhpur, Rajasthan\\
Email: kaur.1@iitj.ac.in}
\and
\IEEEauthorblockN{Atul Kumar}
Department of Chemistry\\
Indian Institute of Technology\\
Jodhpur, Rajasthan\\
Email: atulk@iitj.ac.in}

\maketitle

\IEEEpeerreviewmaketitle

\begin{abstract}
We analyse the role of degree of entanglement for Vaidman's game in a setting where the players share a set of partially entangled three-qubit states. Our results show that the entangled states combined with quantum strategies may not be always helpful in winning a game as opposed to the classical strategies. We further find the conditions under which quantum strategies are always helpful in achieving higher winning probability in the game in comparison to classical strategies. Moreover, we show that a special class of W states can always be used to win the game using quantum strategies irrespective of the degree of entanglement between the three qubits. Our analysis also helps us in comparing the Vaidman's game with the secret sharing protocol. Furthermore, we propose a new Vaidman-type game where the rule maker itself is entangled with the other two players and acts as a facilitator to share a secret key with the two players. 
\end{abstract}


\section{Introduction}
Game theory is an eminently interesting and flourishing field of study, wherein many situations of conflicts can be efficiently examined and resolved \cite{GameTheory}. With the advent of quantum information and computation, game theory has generated a lot of interest in analysing quantum communication protocols from the perspective of a game \cite{BB84, BB84Game}. The analysis not only allows one to study the fundamental of quantum mechanics but also provides a much better insight to the communication protocol in terms of security, payoffs of different players, and complex nature of multi-qubit entanglement. The aim is to study and compare the payoffs of different users and security of a protocol using classical and quantum strategies. In general, quantum strategies are found to be preferable in comparison to the classical strategies. For example, Meyer demonstrated how quantum strategies can be utilized by a player to defeat his classical opponent in a classical penny flip game \cite{Meyer}. He further explained the relation of penny flip game setting to efficient quantum algorithms. Similarly, Eisert \cite{Eisert} suggested a quantum solution for avoiding the Prisoners' Dilemma. Moreover, the quantum version of Prisoners' Dilemma game was also experimentally realized using a NMR quantum computer \cite{Du}. On the other hand, Anand and Benjamin \cite{Anand} found that for a scenario in penny flip game where two players share an entangled state, a player opting for a mixed strategy can still win against a player opting for a quantum strategy. Therefore, it becomes important to analyse the role of quantum entanglement in game theory. Furthermore, one must also understand and study the importance of using different entangled systems under different game scenarios to take the advantage of usefulness of such entangled systems in different situations.   \par
In this article, we analyse a game proposed by Vaidman \cite{Vaidman} in which a team of three players always wins the game, when they share a three qubit maximally entangled state. The team, however, does not win the game when players opt for pure classical strategies, in fact the maximum winning probability that can be achieved using classical strategies is $3/4$. Our analysis of Vaidman game includes two different classes of three-qubit entangled states, namely, GHZ class \cite{GHZpaper} and W class of states \cite{Dur}. We attempt to establish a             
relation between the winning probability of Vaidman's game \cite{Vaidman} with the degree of entanglement of various three-qubit entangled states used as a resources in the game. Interestingly, our results show that for GHZ class, there are set of states for which classical strategies give better winning probability than the quantum strategies. In comparison to the GHZ class, for a special class of W states, quantum strategies prove to be always better than the classical strategies. We further establish a direct correspondence between  Vaidman's game and Quantum Secret Sharing (QSS) \cite{Hillery}. In addition, we also propose a Vaidman-type game where one of the players sharing the three-qubit entanglement defines the rule of a game to be played between him/her and the other two players. A detailed examination of the proposed game shows that  the rule-maker finds himself in an advantageous situation whenever they share a partially entangled state, because this enables the rule-maker to modify rules in such a way that the team of other two players loose the game. Moreover, we further suggest an application of such a game in facilitated secret sharing between three parties, where one of the players is a facilitator and also controls the secret sharing protocol. \par
The organization of the article is as follows. In Section II and III, we briefly describe three-party entanglement (to identify GHZ and W class of states) and QSS, respectively. In section IV, we establish a correspondence of Vaidman's game with QSS. In corresponding subsections we further demonstrate the outcomes of using GHZ and W class of states for Vaidman's game. A new Vaidman-type game is proposed in the Section V followed by its application for QSS in the subsection. We conclude the article in the Section VI.

\section{Three-qubit Entanglement}
Dur et al. \cite{Dur} classified pure states of a three-qubit entangled systems in two inequivalent classes, namely GHZ class and W class represented as
\begin{equation}
\label{GHZgeneral}
\vert{\psi_{GHZ}}\rangle= sin\theta \vert{000}\rangle + cos\theta \vert{111}\rangle
\end{equation}
and
\begin{equation}
\label{Wgeneral}
\vert{\psi_{W}}\rangle= a\vert{100}\rangle + b\vert{010}\rangle + c\vert{001}\rangle, 
\end{equation}
respectively where $\theta\in\left( 0,\pi/4\right)$ and $\left| a \right|^2  + \left| b \right|^2  + \left| c \right|^2  = 1$. The above two classes are termed as inequivalent classes as a state belonging to one of the class states cannot be converted to a state belonging to another class by performing any number of Local Operations and Classical Communication (LOCC). The degree of entanglement for a pure three-qubit system can be defined using a measure called three-tangle $(\tau)$ \cite{Coffman}, given by 
\begin{equation}
\label{tau_formula}
\tau= C^2_{P(QR)}-C^2_{PQ}-C^2_{PR}
\end{equation}
where $C_{P(QR)}$ represents the concurrence of the qubit P, with qubits Q and R taken together as one entity \cite{Hill, Wootters1, Wootters2}. The terms $C_{PQ}$ and $C_{PR}$ can be defined in a similar fashion such that,
\begin{equation}
\label{conc_formula}
C(\vert{\psi}\rangle)=\vert\langle\psi\vert\sigma_y\otimes\sigma_y\vert\psi^*\rangle\vert
\end{equation}
Here, $\psi^*$ denotes the complex conjugate of the wave function representing the two-qubit entangled state. The value of three-tangle varies between 0 for product states to 1 for states having maximum entanglement. For example, the three-tangle for maximally entangled GHZ states represented as   
\begin{equation}
\label{GHZ}
\vert{GHZ}\rangle= \frac{1}{\sqrt{2}}(\vert{000}\rangle+\vert{111}\rangle)
\end{equation}
is 1. Similarly the standard state in W class is represented by 
\begin{equation}
\label{W}
\vert{W}\rangle= \frac{1}{\sqrt{3}}(\vert{001}\rangle+\vert{010}\rangle+\vert{001}\rangle)
\end{equation}
Although the standard $W$ state possesses genuine three-qubit entanglement, the same cannot be identified using the three-tangle as an entanglement measure as the three-tangle of the standard $W$ state is 0. Nevertheless, one can be assured that the W class of states exhibit genuine tripartite entanglement using other entanglement measures such as the sum of the concurrences for all the three bipartite entities, i.e., $C_{AB}+C_{BC}+C_{CA}$. The maximum value of sum of three concurrences is 2, which is attained for a standard $W$ state, as shown in (\ref{W}). 
\section{Quantum Secret Sharing}
Secret sharing is the process of splitting a secret message into parts, such that no part of it is sufficient to retrieve the original message \cite{Hillery}. The original idea was to split the information between the two recipients, one of which may be a cheat (unknown to the sender). Only when the two recipients cooperate with each other, they retrieve the original message. The protocol, therefore, assumes that the honest recipient will not allow the dishonest recipient to cheat, hence, splitting the information between the two. \par
The original protocol can be implemented using the maximally entangled three-qubit GHZ state, as given in (\ref{GHZ}), shared between the three users Alice, Bob, and Charlie. Alice splits the original information between Bob and Charlie in a way that the complete message cannot be recovered unless they cooperate with each other. For sharing a joint key with Bob and Charlie, Alice suggests all of them to measure their qubits either in X or Y direction at random where the eigen states in X and Y basis are defined as
\begin{equation}
\label{basis}
\vert{\pm x}\rangle=\frac{1}{\sqrt{2}}(\vert{0}\rangle \pm \vert{1}\rangle), \ \ \ \vert{\pm y}\rangle=\frac{1}{\sqrt{2}}(\vert{0}\rangle \pm i\vert{1}\rangle)
\end{equation}
\begin{table}[!t !h]
\renewcommand{\arraystretch}{1.3}
\caption{Effect of Bob's and Charlie's measurement on Alice's state in a GHZ state}
\label{QSS}
\centering
\begin{tabular}{|c|cccc|}
\hline
\backslashbox{Bob}{Charlie} & $\vert{+x}\rangle$ & $\vert{-x}\rangle$ & $\vert{+y}\rangle$ & $\vert{-y}\rangle$ \\
\hline
$\vert{+x}\rangle$ & $\vert{+x}\rangle$ & $\vert{-x}\rangle$ & $\vert{-y}\rangle$ & $\vert{+y}\rangle$ \\
$\vert{-x}\rangle$ & $\vert{-x}\rangle$ & $\vert{+x}\rangle$ & $\vert{+y}\rangle$ & $\vert{-y}\rangle$ \\
$\vert{+y}\rangle$ & $\vert{-y}\rangle$ & $\vert{+y}\rangle$ & $\vert{-x}\rangle$ & $\vert{+x}\rangle$ \\
$\vert{-y}\rangle$ & $\vert{+y}\rangle$ & $\vert{-y}\rangle$ & $\vert{+x}\rangle$ & $\vert{-x}\rangle$ \\
\hline
\end{tabular}
\end{table}
The effects of Bob's and Charlie's measurements on the state of Alice's qubit is shown in Table \ref{QSS}. After performing their measurements at random, Bob and Charlie announce their choice of measurement bases (but not the measurement outcomes) to Alice. This is followed by Alice telling her choice of measurement bases to Bob and Charlie. Only the bases XXX, XYY, YXY, and YYX (for Alice, Bob, and Charlie, respectively) are accepted, for sharing the secret key. The results from all other random choices of bases are discarded. 

Bob and Charlie must meet and tell each other their measurement outcomes so as to collectively know the measurement outcome of Alice. For instance, if both Bob and Charlie measure in X basis and their measurement outcomes are $+1$ and $+1$ respectively, or $-1$ and $-1$ respectively then the corresponding outcome of Alice will be $+1$ when measured in X basis. On the other hand, if the measurement outcomes of Bob and Charlie are $+1$ and $-1$ respectively or vice-versa, then the corresponding outcome of Alice will be $-1$ when measured in X basis.

\section{Vaidman's Game representing Quantum Secret Sharing}
In this section, we show a correspondence between the QSS protocol \cite{Hillery} to the Vaidman's game \cite{Vaidman}. We, therefore, first briefly describe the Vaidman's game. In this game, three players, namely Alice, Bob, and Charlie, are taken to arbitrary remote locations: A, B, and C, respectively. Now each player is asked one of the two possible questions: Either \textquotedblleft{What is X?}\textquotedblright \ or \textquotedblleft{What is Y?}\textquotedblright . The players can give only two possible answers, either -1 or +1. The rules of the game suggest that either each player is asked the X question or two of the players are asked the Y question and the remaining one is asked the X question. The team of three players wins the game if the product of their answers is +1 (when all are asked the X question) and -1 (when one is asked the X question and two are asked the Y question). Clearly, if the players adopt the classical strategy then at best they can achieve a winning probability of $3/4$. On the other hand, if the three players share a three-qubit maximally entangled GHZ state, as shown in (\ref{GHZ}), then they always win the game by using a simple quantum strategy, i.e., whenever a player is asked the X(Y) question, she/he measures her/his qubit in the X(Y) basis and uses the measurement outcome obtained in the measurement process as her/his answer.
\subsection{Use of GHZ class states}
That the three players always win the game using the above strategy is because of the strong correlations between the three qubits of the GHZ state. For example, the three qubits in the GHZ state are related as
\begin{eqnarray}
\label{Vaidman Game}
\lbrace{M^X_A}\rbrace\lbrace{M^X_B}\rbrace\lbrace{M^X_C}\rbrace &=&1  \IEEEnonumber \\
\lbrace{M^X_A}\rbrace\lbrace{M^Y_B}\rbrace\lbrace{M^Y_C}\rbrace &=&-1 \IEEEnonumber \\
\lbrace{M^Y_A}\rbrace\lbrace{M^X_B}\rbrace\lbrace{M^Y_C}\rbrace &=&-1 \IEEEnonumber \\
\lbrace{M^Y_A}\rbrace\lbrace{M^Y_B}\rbrace\lbrace{M^X_C}\rbrace &=&-1 
\end{eqnarray}
where $\lbrace{M^X_i}\rbrace$ is the measurement outcome of the $'i'$-th player  measuring her/his qubit in X basis, and $\lbrace{M^Y_i}\rbrace$ is the measurement outcome of the $'i'$-th player measuring her/his qubit in Y basis. A clear correspondence between the Vaidman's game and the QSS protocol is shown in (\ref{Vaidman Game}). \par
\begin{figure}[!t]
\centering
\includegraphics[width=2.5in]{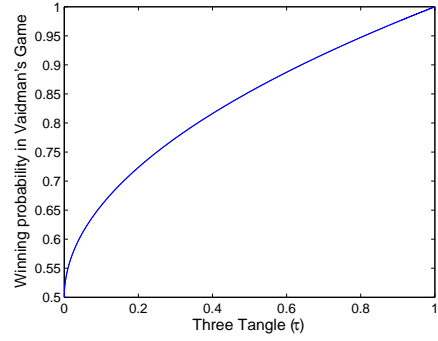}
\caption{Success probability of winning Vaidman's game using GHZ-type states}
\label{fig_genGHZ_VaidmanGame}
\end{figure}

We now proceed to analyze the Vaidman's game in a more general setting where the three players share a general GHZ state represented in (\ref{GHZgeneral}), instead of sharing a maximally entangled GHZ state as described in the original game. Clearly, for a general GHZ state, the success probability of winning the above defined game varies from $50\%$ to $100\%$ as shown in Figure \ref{fig_genGHZ_VaidmanGame}. Here, we have assumed that the probability of players being asked the set of 4 questions ($XXX$, $XYY$, $YXY$, $YYX$) is equally likely.
In Figure \ref{fig_genGHZ_VaidmanGame}, the winning probability of the game, i.e, $\frac{1}{2}(1+sin2\theta)$ is plotted against the degree of entanglement, three tangle ($\tau$). It is clear that only for maximally entangled state, i.e., when $\tau$ attains its maximum value (at $\theta=\pi/4$), the players have $100\%$ chances of winning the game. For all other values of $\tau$ the success probability is always less than the one obtained with a maximally entangled state. Interestingly, only the set of states with $\tau >0.25$ achieve better success probability in comparison to a situation where all the three players opt for classical strategies. Therefore, for the set of states with $\tau <0.25$, classical strategies will prove to be better in comparison to quantum strategies. Hence, entanglement may not be always useful in winning the games using quantum strategies in comparison to classical strategies. 

\subsection{Use of W class states}
Although W-type states belong to a different class of states, they can also be used as a resource in winning Vaidman's game with a different set of questions. In this case, the playes may be asked either \textquotedblleft{What is Z?}\textquotedblright \ or \textquotedblleft{What is Y?}\textquotedblright. The answers to these questions can again be either +1 or -1. For this, either all players are asked the Z question; or one of the players is asked the Z question and the remaining are asked the Y question. The players win the game if the product of their answers is -1, if all are asked the Z question; and +1, in all other cases. If the players share the standard $W$ state, given in (\ref{W}), before the start of play then they can win this game with a success probability of $87.5\%$. On similar grounds, we can use the standard $W$ state for probabilistic QSS, as QSS holds direct correspondence with the Vaidman's game. \par
\begin{figure}[!t]
\centering
\includegraphics[width=2.5in]{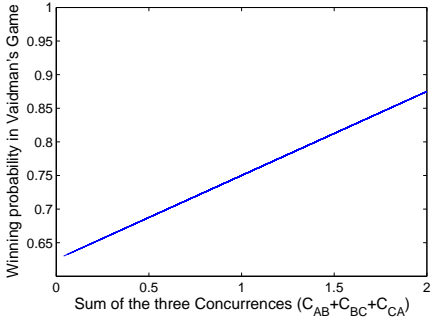}
\caption{Success probability of winning Vaidman's game using W-type states}
\label{fig_genW_VaidmanGame1}
\end{figure}

Similar to the case of GHZ class, here, we analyze the success probability of the Vaidman's game if the three players share a general W-type state as shown in (\ref{Wgeneral}). In such a scenario, the team wins the game with a success probability given as $\frac{1}{4}(\frac{5}{2}+bc+ab+ac)$. This value holds true for an assumption that the team will be asked the 4 set of questions ($ZZZ$, $ZYY$, $YZY$, $YYZ$) with equal likelihood. The plot of winning probability of Vaidman's game versus the sum of three concurrences is demonstrated in Figure \ref{fig_genW_VaidmanGame1}. The figure shows that the winning probability of Vaidman's game linearly increase with the sum of concurrences for W-type states. Furthermore, the plot also indicates that for W-type states shown in (\ref{Wgeneral}) with sum of two qubit concurrences exceeding 1, the winning probability of Vaidman's game is always greater than the classical winning probability. Also, the highest success probability of $87.5\%$ can be achieved for values $a=b=c=\frac{1}{\sqrt{3}}$. \par
Although the use of partially entangled systems, in general, leads to probabilistic information transfer \cite{Karlsson, Shi}, Pati and Agrawal \cite{Agrawal} have shown that there exists a special class of W-type states which can be used for perfect teleportation and dense coding. The class of states can be represented as 
\begin{equation}
\label{Wn}
\vert{W_n}\rangle= \frac{1}{\sqrt{2(1+n)}}(\vert{100}\rangle+\sqrt{n}e^{i\gamma}\vert{010}\rangle+\sqrt{n+1}e^{i\delta}\vert{001}\rangle)
\end{equation}
\begin{figure}[!t]
\centering
\includegraphics[width=2.5in]{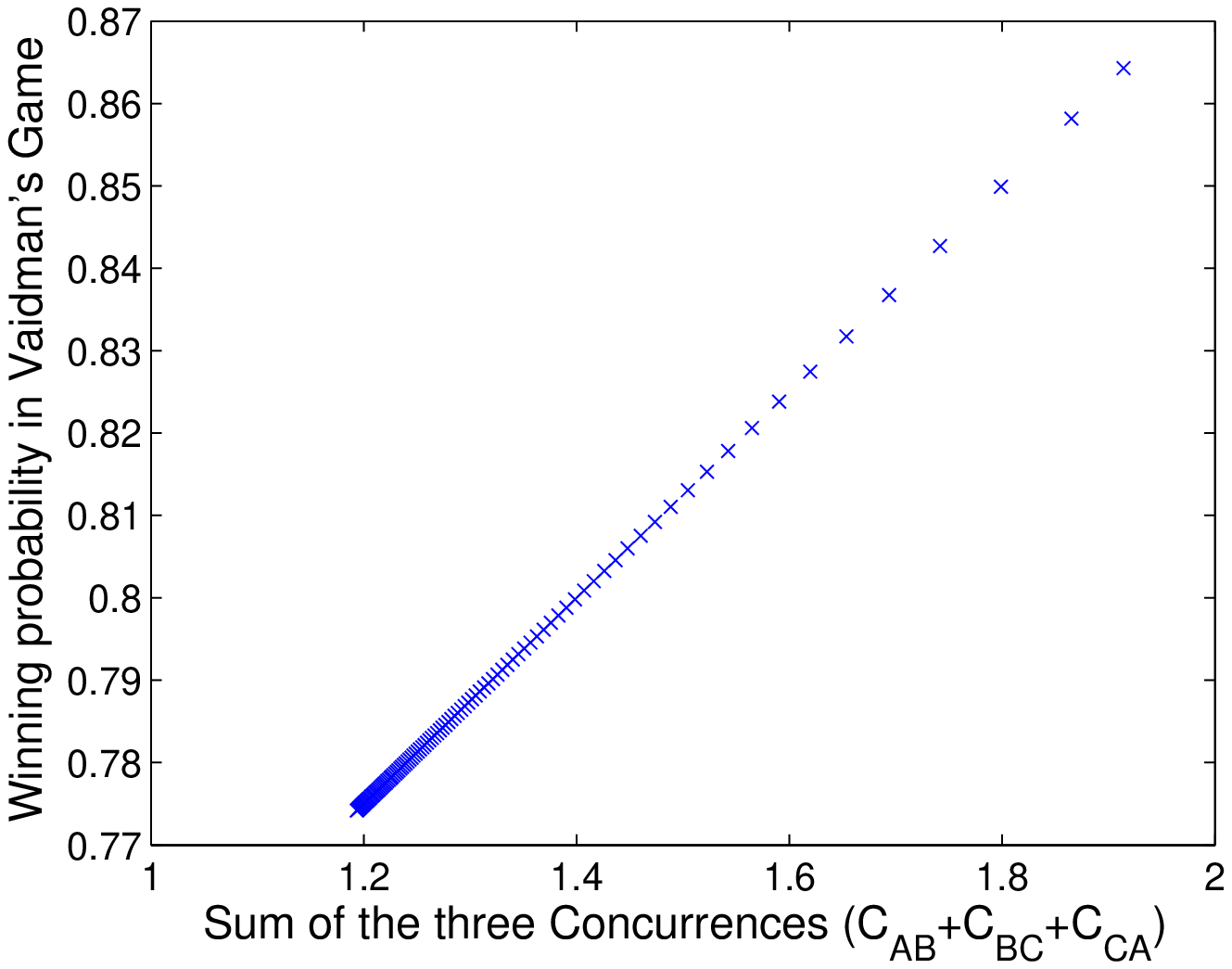}
\caption{Success probability of winning Vaidman's game using $W_n$ states}
\label{fig_Wn_VaidmanGame1}
\end{figure}
\begin{figure}[!t]
\centering
\includegraphics[width=2.5in]{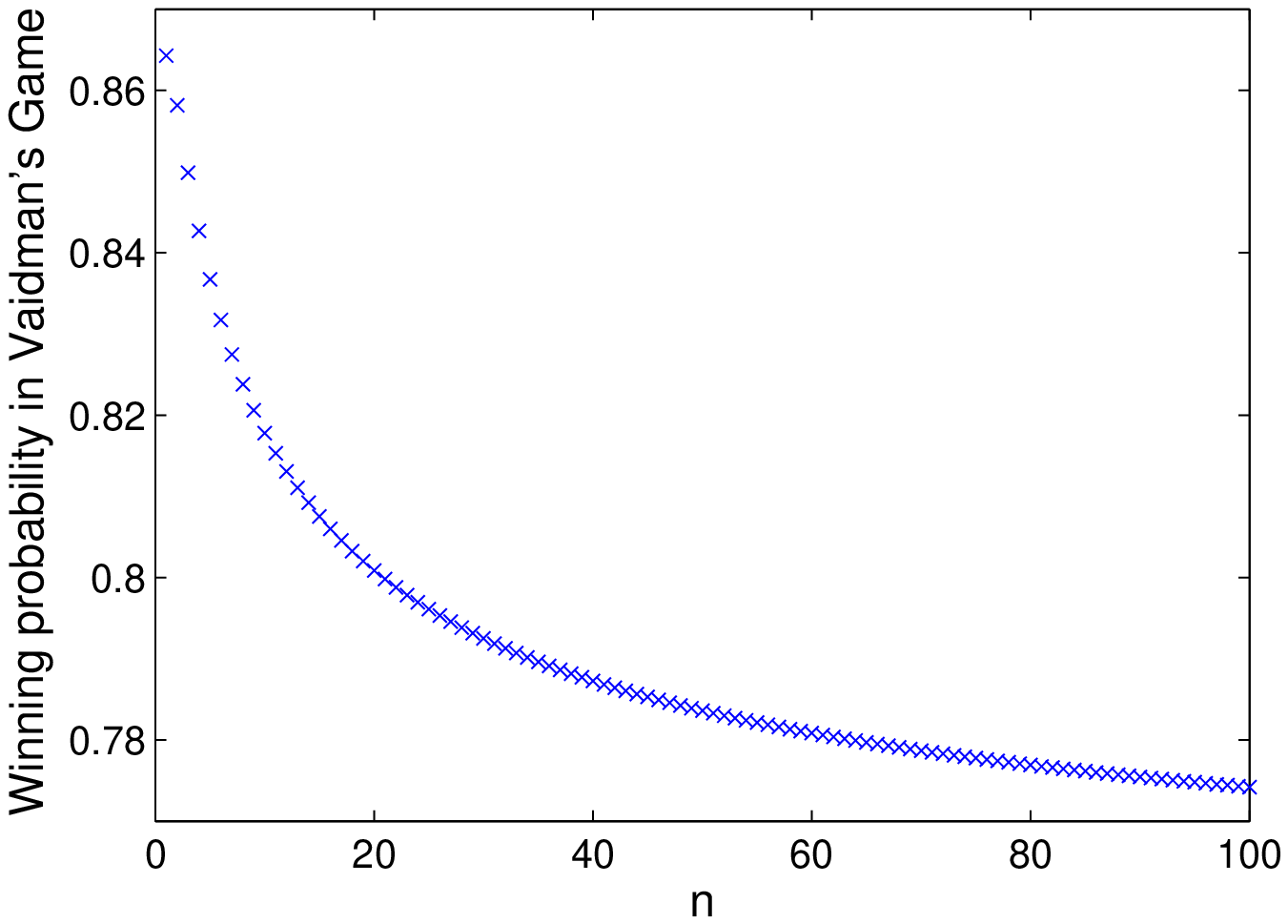}
\caption{Success probability of winning Vaidman's game using $W_n$ states}
\label{fig_Wn_VaidmanGame2}
\end{figure}
where $n$ is a positive integer and $\delta$ and $\gamma$ are relative phases. This motivates us to analyse the usefulness of these states for the Vaidman's game. The success probability that can be obtained by sharing $W_n$ states between the three players can be given by  $\frac{1}{8(n+1)}(5+5n+\sqrt{n+1}+\sqrt{n}(\sqrt{n+1}+1))$. Figure \ref{fig_Wn_VaidmanGame1} clearly demonstrates that if the three players share $W_n$ states, then the success probability using quantum strategies is always greater than the success probability using the classical strategies, independent of the value of the sum of concurrences. Moreover, Figure \ref{fig_Wn_VaidmanGame2} depicts the dependence of winning probability of Vaidman's game on parameter $n$. The highest success probability of $0.86425$ is achieved for $n=1$ when the sum of the three concurrences is $1.914$. Nevertheless, the winning probability is always greater than the one obtained using classical strategies. 

\subsection{Comparison of the use of GHZ and W states}
The above calculations suggest that although a standard GHZ state achieves $100\%$ success probability in winning the Vaidman's game which is more than the winning probability achieved by the standard $W$ state, only the set of GHZ-type states with a value of $\tau>0.25$ are useful for obtaining the success probability greater than the one obtained using classical strategies. Moreover, all the W-type states with the sum of three concurrences greater than one, can be useful in winning the game. In addition, a special class of W-type states, i.e., $W_n$ states give better prospects of winning the Vaidman's game, than any classical means, for all values of $n$. 

\section{A game where the rule-maker is entangled with the players}
The essence of Vaidman's game can be efficiently employed in an interesting scenario, where the rule-maker itself is entangled with the players playing the Vaidman-type game. In our proposed game, Alice, Bob and Charlie share a three-qubit entangled state. We assume that Charlie  prepares a three-qubit state and gives one qubit each to Alice ($A$) and Bob ($B$), keeping one ($C$) qubit with himself. Charlie agrees to help Alice and Bob, if they win the game as per the rules defined by Charlie. For this, Charlie measures his qubit in a general basis as shown in (\ref{Parametrized_basis}). Charlie, then asks questions \textquotedblleft{What is X?}\textquotedblright \ or \textquotedblleft{What is Z?}\textquotedblright \ to the team. Alice and Bob are not allowed to discuss and have to give individual answers each. Their answer can be +1 or -1.  If the team is asked the X (Z) question, both Alice and Bob measure their qubits in X (Z) basis and give their measurement results as answers to the asked questions. 
\begin{equation}
\label{Parametrized_basis}
\vert{b_0}\rangle= sin\lambda \vert{0}\rangle - cos\lambda \vert{1}\rangle; \ \ \ \ \
\vert{b_1}\rangle= cos\lambda \vert{0}\rangle + sin\lambda \vert{1}\rangle
\end{equation}

If Charlie's measurement outcome is $\vert{b_0}\rangle$, he declares the winning condition to be as shown in (\ref{rule1}), and if his  measurement outcome is $\vert{b_1}\rangle$, he declares the winning condition to be as shown in (\ref{rule2}). Here, $\lbrace{M^X_i}\rbrace$ is the measurement outcome when the $'i'$-th player  measures her/his qubit in X basis, and $\lbrace{M^Z_i}\rbrace$ is the measurement outcome when the $'i'$-th player measures her/his qubit in Z basis.

\begin{equation}
\label{rule1}
\lbrace{M^X_A}\rbrace\lbrace{M^X_B}\rbrace=1 \ \ \ \ \ \lbrace{M^Z_A}\rbrace\lbrace{M^Z_B}\rbrace=-1
\end{equation}
\begin{equation}
\label{rule2}
\lbrace{M^X_A}\rbrace\lbrace{M^X_B}\rbrace=-1 \ \ \ \ \ \lbrace{M^Z_A}\rbrace\lbrace{M^Z_B}\rbrace=1
\end{equation}

\begin{figure}[!t]
\centering
\includegraphics[width=2.5in]{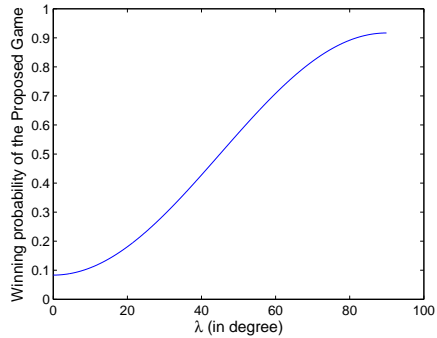}
\caption{Success probability of winning the proposed game where the rule-maker is entangled with the players using a standard $W$ states}
\label{fig_ProposedGame}
\end{figure}

If Charlie prepares a three-qubit GHZ state as shown in (\ref{GHZ}), then the team has $50\%$ success probability irrespective of the measurement basis used by Charlie. However, if Charlie prepares a three-qubit $W$ state as shown in (\ref{W}), then success probability of the team depends on the parameter $\lambda$- governing the basis in which Charlie makes a measurement. A plot of success probability achieved with respect to the parameter $\lambda$ is shown in Figure \ref{fig_ProposedGame}. The maximum winning probability of the team $0.9167$ when $\lambda=90^\circ$, i.e., Charlie measures in computational basis ($\vert{b_0}\rangle=\vert{0}\rangle$ and $\vert{b_1}\rangle=\vert{1}\rangle$). On the other hand, Charlie can also measure in computational basis with $\vert{b_0}\rangle=\vert{1}\rangle$ and $\vert{b_1}\rangle=\vert{0}\rangle$ when $\lambda=0^\circ$. In such a parameter setting, the team mostly looses the game as the winning probability is only $0.0833$. Thus, if Charlie wants to help Alice and Bob, he prefers to prepare a standard $W$ state and performs measurement in the computational basis ($\vert{b_0}\rangle=\vert{0}\rangle$ and $\vert{b_1}\rangle=\vert{1}\rangle$) so that the team can win the game with a success rate of $91.667\%$. In this situation, the use of quantum strategy is always preferable for the team of Alice and Bob.  

For all the above values and winning probability, we are assuming that Charlie asks the X and Z question with equal probability. Alice and Bob may choose not to measure their qubits and randomly answer as +1 or -1 (classical approach). In that case, the team may win the game half the times. 

\subsection{An application of the above game in secret sharing}
For establishing a relation between the proposed game and secret sharing, we consider that Alice and Bob are kept in two different cells and are partially disallowed to communicate. By partially, we mean that they can can communicate only under the presence of a facilitator or a controller (Charlie in our case), who listens and allows secure communication between the two. To accomplish this task, we prefer to exploit the properties of a standard $W$ state over the use of a $W_1$ state, because the success rate of winning Vaidman's game is $87.5\%$ when a standard $W$ state is shared, as opposed to $86.425\%$ when a $W_1$ state is shared within the team members. Also, we further consider that Charlie performs his measurement in the basis as shown in (\ref{Parametrized_basis}) at $\lambda=90^\circ$, i.e., Charlie measures his qubit in computational basis ($\vert{b_0}\rangle=\vert{0}\rangle$ and $\vert{b_1}\rangle=\vert{1}\rangle$). \par

\begin{table}[t]
\renewcommand{\arraystretch}{1.3}
\caption{Control mode of facilitated information sharing}
\label{table_CM}
\centering
\begin{tabular}{|c|cccccc|}
\hline
Charlie's measurement outcome & $\vert{1}\rangle$ & $\vert{1}\rangle$ & $\vert{1}\rangle$ & $\vert{1}\rangle$ & $\vert{1}\rangle$ & $\vert{1}\rangle$ \\
\hline
Alice's basis & Z & Z & X & X & X & X \\
Bob's basis & Z & X & Z & X & X & X \\
Is the choice of basis accepted? & yes & no & no & yes & yes & yes \\
Alice's measurement outcome & $+1$ & - & - & $+1$ & $-1$ & $+1$ \\
Bob's measurement outcome & $+1$ & - & - & $+1$ & $+1$ & $-1$ \\
\hline
Correlation as expected? & $\checkmark$ & - & - & $\times$ & $\checkmark$ & $\checkmark$ \\
\hline
\multicolumn{7}{|c|}{Alice and Bob are asked to announce their outcome and it is checked if} \\
\multicolumn{7}{|c|}{their results comply with (12) in more than or equal to $75\%$ cases} \\ 
\hline
\end{tabular}
\end{table}

In order to share a key, Charlie chooses to operate in two different modes, namely control mode and message mode. The control mode corresponds to Charlie's measurement outcome $\vert{1}\rangle$, and is used to check whether Alice and Bob are honest or not, as shown in Table \ref{table_CM}. Similarly, the message mode corresponds to Charlie's measurement outcome $\vert{0}\rangle$, and is used to share a secret key with Alice and Bob (Table \ref{table_MM}). For this, Charlie prepares $'m'$ standard $W$ states as shown in (\ref{W}) and distributes qubits 1 and 2 of each state to Alice and Bob, respectively keeping the third qubit with himself. Charlie, then performs a measurement on his qubit in the computational ($\vert{0}\rangle$, $\vert{1}\rangle$) basis. Meanwhile, Alice and Bob randomly choose their basis of measurement (either X or Z) and announce their choice of basis to Charlie. If they choose two different basis, then their choices are discarded. Alternately, Charlie randomly chooses a basis of measurement and announces his choice to Alice and Bob. This will ensure that both Alice and Bob perform measurements in the same basis. This step is repeated for $'m'$ qubits, and Alice and Bob note down their measurement results each time. \par

\begin{table}[!t]
\renewcommand{\arraystretch}{1.3}
\caption{Message mode of facilitated information sharing}
\label{table_MM}
\centering
\begin{tabular}{|c|cccccc|}
\hline
Charlie's measurement outcome & $\vert{0}\rangle$ & $\vert{0}\rangle$ & $\vert{0}\rangle$ & $\vert{0}\rangle$ & $\vert{0}\rangle$ & $\vert{0}\rangle$ \\
\hline
Alice's basis choice & X & X & X & Z & Z & Z \\
Bob's basis choice & X & X & Z & X & Z & Z \\
Basis choice accepted? & yes & yes & no & no & yes & yes \\
Alice's measurement outcome & $\vert{+}\rangle$ & $\vert{-}\rangle$ & - & - & $\vert{0}\rangle$ & $\vert{1}\rangle$ \\
Bob's measurement outcome & $\vert{+}\rangle$ & $\vert{-}\rangle$ & - & - & $\vert{1}\rangle$ & $\vert{0}\rangle$ \\
\hline
\multicolumn{7}{|c|}{$\vert{0}\rangle$ and $\vert{+}\rangle$ correspond to secret bit: 0} \\ 
\multicolumn{7}{|c|}{$\vert{1}\rangle$ and $\vert{-}\rangle$ correspond to secret bit: 1} \\
\hline
\multicolumn{7}{|c|}{Let Charlie announce that Bob should flip his outcome whenever he} \\ 
\multicolumn{7}{|c|}{chooses Z basis for measurement} \\
\hline
Shared secret bit & 0 & 1 & - & - & 0 & 1 \\
\hline
\end{tabular}
\end{table}

If Charlie gets $\vert{0}\rangle$ as his measurement outcome, then he knows that the measurement results of Alice and Bob are related as in (\ref{rule1}) with certainty. As explained above, this will be the message mode of the proposed secret sharing scheme, wherein Alice's and Bob's outcomes will either be same or different. The relation between their outcomes is only known to Charlie, which he announces at the end of the protocol. On the other hand, if Charlie gets $\vert{1}\rangle$ as the measurement outcome, then the measurement results of Alice and Bob are related as in (\ref{rule2}) in $75\%$ cases. Since this is a control mode, Charlie secretly asks both Alice and Bob to announce their measurement outcomes, which he verifies to check if anyone (Alice or Bob) is cheating. If the results announced by Alice and Bob do not comply with the results in (\ref{rule2}) more than $75\%$ times, then cheating is suspected. Moreover, as Alice and Bob are not allowed to discuss, they cannot distinguish between the message and the control mode. If both, Alice and Bob are asked to announce their measurement outcomes, then the control mode of secret sharing is taking place. While, if none of them is asked to announce her/his results, then the message mode of secret sharing occurs. If Charlie suspects cheating in the control mode, he disallows communication and does not announce the relation between the outcomes of Alice and Bob for message runs. However, if Charlie does not find anything suspicious, he announces in the end, which results correspond to message and control mode, and also the relation between the outcomes of Alice's and Bob's measurement outcomes in the message mode. This protocol, therefore, enables the controller to check a pair of agents for their honesty, and simultaneous sharing of a secret key with them, if they are proved honest. 

\section{Conclusion and Future scope}
In this article, we addressed the role of degree of entanglement for Vaidman's game. We analysed the relation between the success probability of the Vaidman's game with the three-qubit entanglement measures considering both quantum and classical strategies. The results obtained here indicate that entanglement and quantum strategies may not be always useful in winning the game. For example, we found that there are set of GHZ class and W class states, for which classical strategies are proved to be better than the quantum strategies. On the other hand, for the special class of W-type states, i.e., $W_{n}$ states, quantum strategies are always better than the classical strategies in winning the Vaidman's game. We further explored a correspondence between the Vaidman's game using general three-qubit pure states and the QSS protocol. In addition, we have proposed an efficient game, where the player deciding the rules of the game itself is entangled with other two players. The proposed game may find an application in facilitated secret sharing, where a facilitator checks the players involved for their honesty and simultaneously controls the process of sharing information between them. \par 
It will be interesting to analyse these games under real situations, i.e., considering the success probability  of the game under noisy conditions. For future study, we also wish to extend our analysis for similar games between four players and then to generalise it for the involvement of $N$ players.

\end{document}